\documentclass[conference]{IEEEtran}
\IEEEoverridecommandlockouts
\usepackage{cite}
\usepackage{amsmath,amssymb,amsfonts}
\usepackage{algorithmic}
\usepackage{graphicx}
\usepackage{textcomp}
\usepackage{xcolor}
\usepackage{multirow} 
\usepackage{adjustbox}
\usepackage{amsmath}
\usepackage{xcolor} 
\usepackage{url} 
\usepackage[normalem]{ulem} 
 
\usepackage{colortbl} 
\def\BibTeX{{\rm B\kern-.05em{\sc i\kern-.025em b}\kern-.08em
    T\kern-.1667em\lower.7ex\hbox{E}\kern-.125emX}}

\begin{document}
    
\title{Pyramid Hierarchical Masked Diffusion Model for Imaging Synthesis}

\author{\IEEEauthorblockN{1\textsuperscript{st} Xiaojiao Xiao}
\IEEEauthorblockA{\textit{Department of Computer Science)} \\
\textit{Toronto Metropolitan University)}\\
Toronto, Canada \\
xiaojiao@torontomu.ca}
\and
\IEEEauthorblockN{2\textsuperscript{nd} Qinmin Vivian Hu}
\IEEEauthorblockA{\textit{Department of Computer Science)} \\
\textit{Toronto Metropolitan University)}\\
Toronto, Canada \\
vivian@torontomu.ca}
\and
\IEEEauthorblockN{3\textsuperscript{rd} Guanghui Wang}
\IEEEauthorblockA{\textit{Department of Computer Science)} \\
\textit{Toronto Metropolitan University)}\\
Toronto, Canada \\
wangcs@torontomu.ca}
}

\maketitle
\begin{abstract}
Medical image synthesis plays a crucial role in clinical workflows, addressing the common issue of missing imaging modalities due to factors such as extended scan times, scan corruption, artifacts, patient motion, and intolerance to contrast agents. The paper presents a novel image synthesis network, the Pyramid Hierarchical Masked Diffusion Model (PHMDiff), which employs a multi-scale hierarchical approach for more detailed control over synthesizing high-quality images across different resolutions and layers. Specifically, this model utilizes randomly multi-scale high-proportion masks to speed up diffusion model training, and balances detail fidelity and overall structure. The integration of a Transformer-based Diffusion model process incorporates cross-granularity regularization, modeling the mutual information consistency across each granularity's latent spaces, thereby enhancing pixel-level perceptual accuracy. Comprehensive experiments on two challenging datasets demonstrate that PHMDiff achieves superior performance in both the Peak Signal-to-Noise Ratio (PSNR) and Structural Similarity Index Measure (SSIM), highlighting its capability to produce high-quality synthesized images with excellent structural integrity. Ablation studies further confirm the contributions of each component. Furthermore, the PHMDiff model, a multi-scale image synthesis framework across and within medical imaging modalities, shows significant advantages over other methods. The source code will be released with the paper. The source code is available at \url{https://github.com/xiaojiao929/PHMDiff}
\end{abstract}

\section{Introduction}

Medical image synthesis across and within medical imaging modalities plays a crucial role in optimizing clinical workflows, especially in the high-demand fields of radiology and radiation oncology \cite{wang2021review}. Different modalities, such as CT, MRI, and PET, or variations in spatial resolution (e.g., 3T vs. 7T), often provide complementary information, including detailed anatomical structures and nuanced abnormalities. However, conventional acquisition is frequently unfeasible due to constraints on time, cost, labor, or safety concerns such as radiation exposure. As a result, image synthesis has become a primary method for substituting or expediting imaging procedures without incurring additional costs or risks.

Deep learning-based synthesis techniques have made significant strides in the field of medical imaging, particularly through the use of Generative Adversarial Networks (GANs) \cite{goodfellow2020generative} and their various adaptations, such as DCGAN, WGAN, CGAN, and CycleGAN. These models often struggle with unstable training and mode collapse, which limits the diversity and fidelity of the generated synthetic images \cite{zhang2018convergence,zhang2021six}. To address these limitations, denoising diffusion models, which generate higher-quality and more diverse synthetic images through a simple iterative process of refining noisy samples, are increasingly becoming an alternative to GANs \cite{ho2020denoising,dhariwal2021diffusion}. Moreover, Masked Autoencoders (MAE) \cite{he2022masked} demonstrate strong recognition performance by learning to regress pixels of masked patches given the other visible patches. Inspired by this, we incorporate masking into transformer-based diffusion models, which can enhance generalization capabilities and the acquisition of a comprehensive understanding of the structural characteristics of medical imaging.
\begin{figure}[t]
\center
\includegraphics[width=0.48\textwidth]{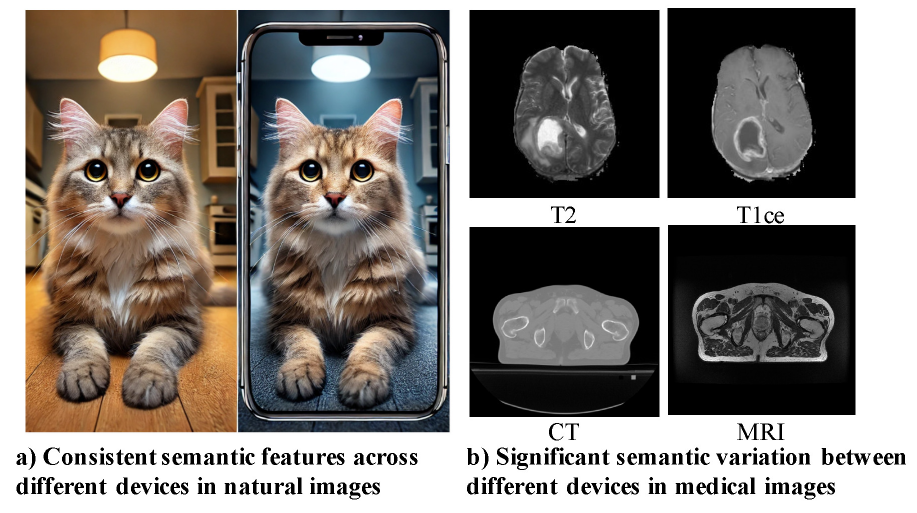}
\caption{Challenge in modeling reliable synthesized medical images.} \label{figure1}
\end{figure}

Despite significant advancements in existing works, several limitations remain: (i) The exclusive use of Mean Squared Error (MSE) loss for reconstruction optimization often results in output images that are blurrier than the original inputs \cite{sajjadi2017enhancenet}. Incorporating a perceptual loss that emphasizes pixel quality could improve fine-grained semantic understanding and representation learning, leading to more realistic synthesized patches. (ii) High rates of random masking can lead to underutilization of images and extended training times. More critically, this approach introduces less reliable features, which undermines the model's generalization capabilities in downstream tasks \cite{he2016deep,he2022masked}. (iii) A further challenge is the inherent appearance of discrepancies between different imaging modalities, which demand extensive modeling efforts, as illustrated in Fig.\ref{figure1}. Moreover, it is crucial for models to perform reliably not only within individual modalities but also across multiple modalities, thereby enhancing their applicability and robustness in multimodal scenarios.

In this research, we introduce a novel image synthesis network named the Pyramid Hierarchical Masked Diffusion Model (PHMDiff), designed to generate high-resolution medical images both across and within different imaging modalities. Our approach begins by decomposing the original image into a multi-resolution pyramid structure, allowing us to capture details and structures at different resolution levels effectively. Starting at the lowest resolution, PHMDiff denoises and reconstructs the image, progressively employing a coarse-to-fine upscaling method to restore and enrich details, ultimately enhancing the overall image quality. At each level of the pyramid, a unique random mask is applied based on the specific resolution and content, leveraging visible parts of the image to guide the reconstruction process. This approach ensures a delicate balance between preserving local details and maintaining overall structural integrity. Then, the processed image is diffused, which speeds up the network training. Additionally, we incorporate a regularization loss to model mutual information across different spatial granularities, optimizing the consistency between pixel-level details and overall structure, which enhances the precision and coherence of the final synthesized image.

Our contributions are summarized as follows:
\begin{itemize}
\item We introduce an innovative pyramid hierarchical masking strategy that balances detail and structure at the image level, effectively preserving crucial fine-grained information.
\item We incorporate cross-granularity regularization (CGR) to model the consistency of mutual information across different granularities, thereby optimizing perceptual accuracy at the pixel level.
\item To our knowledge, this is the first implementation of an end-to-end diffusion model guided by a pyramid hierarchical masking strategy, which has faster training speed and achieves high-quality image synthesis across multiple resolutions and modalities.
\end{itemize}

\section{Related Work}

\subsection{Medical Imaging Synthesis}
Medical imaging synthesis across and within modalities is a critical area of clinical research. This field has witnessed significant advancements with the adoption of deep learning techniques. Early studies employed CNN-based approaches which, while pioneering, often lost intricate structural details due to their reliance on pixel-wise loss functions \cite{isola2017image}. To address these limitations, Generative Adversarial Networks (GANs) were introduced, enhancing the capture of distributional characteristics of target modalities based on source images \cite{goodfellow2020generative}. GANs have shown superior performance across various synthesis tasks, including multi-modal and cross-modality synthesis (e.g., CT to PET, MR to CT), high-resolution conversions (3T-to-7T MRI), and multi-contrast MRI synthesis \cite{dar2019image,ge2019unpaired,xiang2018unpaired,sharma2019missing,yu2019ea,wang2020synthesize,cao2023autoencoder,zhang2023unified,zhang2024unified}. However, GANs often encounter issues with unstable training dynamics and mode collapse, which impact the diversity and fidelity of the synthesized images \cite{zhang2018convergence}.

\subsection{Diffusion Model}
In response to these limitations, Denoising Diffusion Probabilistic Models (DDPMs) have recently emerged as an effective alternative. DDPMs utilize a Markov chain-based process to iteratively refine noisy samples into high-quality synthetic images, thereby progressively improving image quality in generation and synthesis tasks \cite{ho2020denoising,li2022srdiff,gao2023implicit,muller2023multimodal,khader2023denoising}. Despite these successes, most existing diffusion models achieve exceptional performance in sample quality metrics by incorporating complex methodologies, including additional image classifiers. Notably, latent diffusion models \cite{rombach2022high} incorporate self-attention mechanisms \cite{vaswani2017attention}, which facilitate the consideration of context information and the capture of long-distance relationships. Examples include Vtgan \cite{kamran2021vtgan}, GANBERT \cite{shin2020ganbert}, Ptnet \cite{zhang2021ptnet}, and recent advancements in diffusion methods \cite{wei2023diffusion,pan20232d,xiao2024fgc2f}.

\section{The Proposed Method}

\begin{figure}[t]
\center
\includegraphics[width=0.47\textwidth]{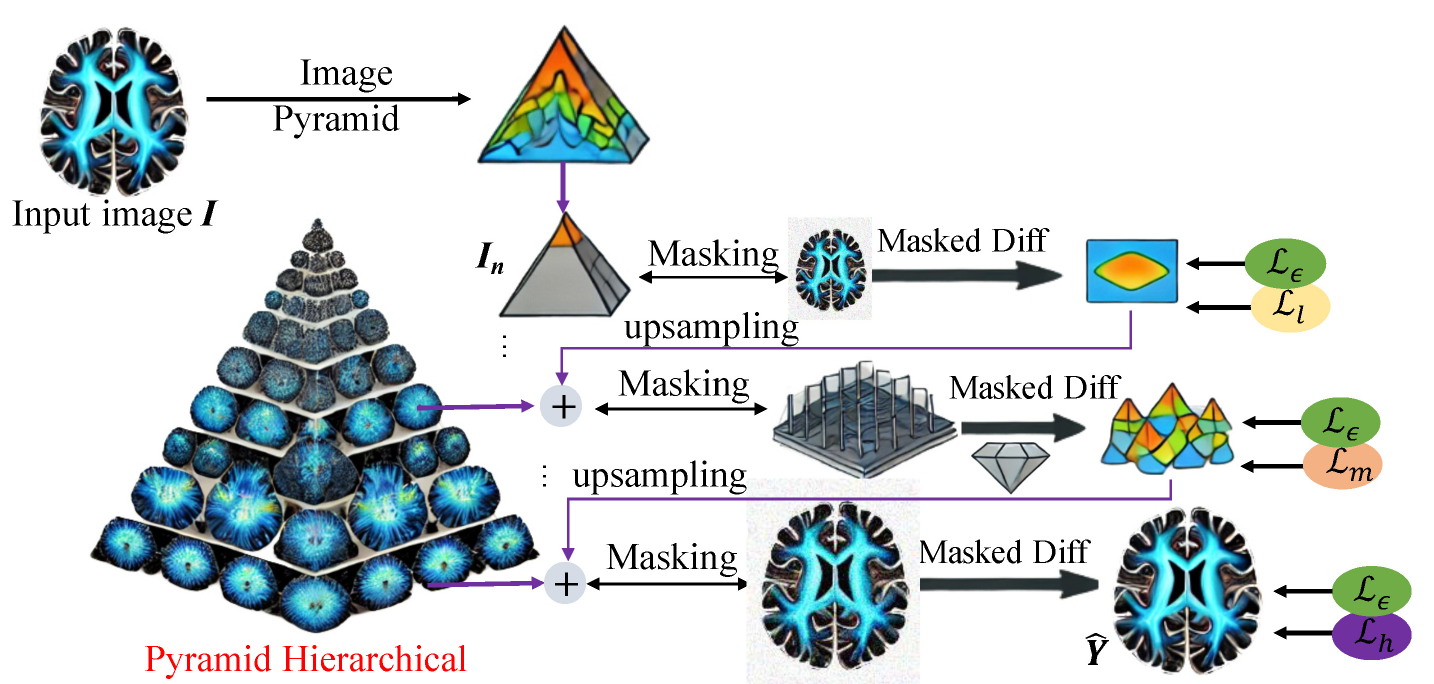}
\caption{Illustration of our proposed framework.} \label{method}
\end{figure}

As depicted in Fig. \ref{method}, our objective is to train the PHMDiff model to synthesize the image $\hat{Y}$ from the input data $I$. Specifically, the input image $I$ is decomposed into multi-scale images to form a pyramid hierarchical coarse-to-fine synthesis. This layered approach ensures the precise capture of structural information at each level of the original image. At each layer, unique masks are generated based on the resolution and content specificity of the image. These masks are designed to obscure specific areas randomly, enabling the Transformer-embedded diffusion model to utilize information from the visible parts of the image for conducting noise addition and reverse processes, thereby capturing global dependencies across the image.

\subsection{Pyramid Hierarchical}\label{ph}
Our PHMDiff approach employs a pyramid hierarchical structure that begins at the lowest resolution and progressively refines upwards to higher levels. This coarse-to-fine approach gradually enhances the richness of image details and effectively utilizes the structural information from previous layers to support finer detail processing at higher levels. Additionally, it allows the model to independently adjust details and structure at different resolution levels, reducing information loss and more accurately maintaining critical anatomical structures and lesion areas. This structure improves image quality and enhances the model's flexibility and efficiency in handling complex images.
Specifically, we decompose the input image $I$, represented in the space $R^{H \times W}$, into a pyramid hierarchical(PH) of multi-scale images, each layer having a progressively lower resolution, where $H$ and $W$ denote the height and width, respectively. At each layer $n$, the image $I_n$ is generated by resizing the image from the previous level, $I_{n-1}$. The dimensions of $I_n$ are calculated as:
\begin{equation} \label{eq1}
W_{n}=\alpha \times W_{n-1}, H_{n}=\alpha \times H_{n-1}
\end{equation}
where $\alpha$ is a scaling factor constrained within $0 < \alpha < 1$; typically, $\alpha$ is set to 0.5, thereby halving the resolution at each step. The output comprises a sequence of images ${I_{0}, I_{1}, \ldots, I_{n}}$, each progressively down-scaled from the preceding one, thus forming the pyramid.

Starting from the lowest resolution $I_n$, the image undergoes progressive denoising and reconstruction at each level, ensuring the accurate capture of the original image’s structural information. As the reconstruction proceeds, the result of each layer is upsampled by a magnification factor corresponding to $\alpha$ and merged with the input of the next layer. This fusion process preserves content consistency, safeguarding against losing essential details that might otherwise occur at lower resolutions. 

\subsection{Architecture design}

\begin{figure}[t]
\center
\includegraphics[width=0.47\textwidth]{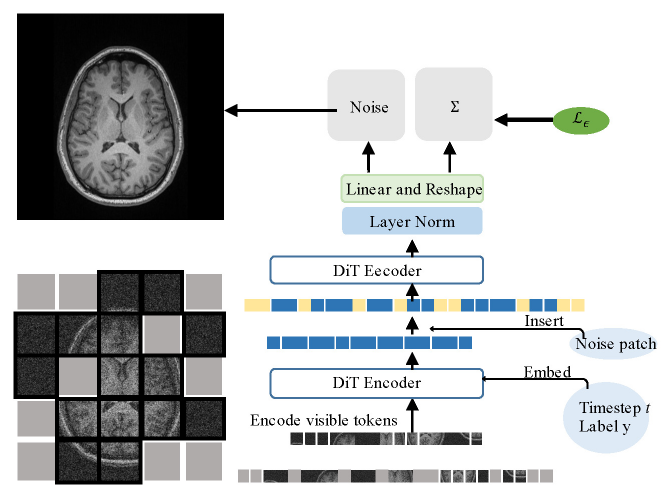}
\caption{Illustration of our proposed masked diff architecture.} \label{masked}
\end{figure}

Our model integrates MAE and DiT \cite{peebles2023scalable} to significantly enhance both the efficiency and effectiveness of the image synthesis process, while concurrently improving the robustness and flexibility in handling complex image data, as shown in Fig.\ref{masked}. The MAE excels at managing local details to maintain visual coherence across the entire image. In contrast, the Diffusion model meticulously adjusts parameters to capture the global structure of the image. Furthermore, we utilize Transformer technology to capture global dependencies throughout the image, thus ensuring both coherence and integrity in the synthesized images. A key component of our methodology is the incorporation of Cross-Granularity Regularization (CGR), which models the consistency of mutual information across various granularities, optimizing perceptual accuracy at the pixel level. 

\subsubsection{Multi-scale Masking}

In our pyramid hierarchical model, we start by diffusing a clean image \(x_0\) with dimensions \(H \times W\) by adding Gaussian noise to create a diffused image \(x_t\) at each timestep \(t\). We then patchify \(x_t\) into \(N\) non-overlapping patches, where \(N\) is determined by \(N = \frac{HW}{p^2}\) for patches of size \(p \times p\). Adaptive masking is applied at each pyramid level, adjusting the masking ratio \(r\) based on the resolution and complexity at that level, and \(\lfloor rN \rfloor\) patches are randomly removed, leaving \(N - \lfloor rN \rfloor\) unmasked patches. These patches are fed into a diffusion model within a multi-resolution pyramid framework, starting from the lowest resolution and progressively processing through finer levels.

\subsubsection{Encoder} In our pyramid hierarchical model, we start by diffusing a clean image \(x_0\) with dimensions \(H \times W\) by adding Gaussian noise to create a diffused image \(x_t\) at each timestep \(t\). We then patchify \(x_t\) into \(N\) non-overlapping patches, where \(N\) is determined by \(N = \frac{HW}{p^2}\) for patches of size \(p \times p\). Adaptive masking is applied at each pyramid level, adjusting the masking ratio \(r\) based on the resolution and complexity at that level, and \(\lfloor rN \rfloor\) patches are randomly removed, leaving \(N - \lfloor rN \rfloor\) unmasked patches. 

\subsubsection{Decoder} The encoder utilizes a standard Vision Transformer (ViT). For instance, consider a training sample $x_{i}$, represented as $x_{i} \sim p(x_{i})$. PHMDiff spatially divides $x_{t}$ into two non-overlapping regions: the masked region $x_{t}^{m}$ and the visible region $x_{t}^{v}$. The ViT encoder $E_\varphi(\cdot)$ processes only the visible patches $x_{0}^{v}$, encoding each patch into the latent space. The output from this encoding, $E_\varphi(x_{t}^{v})$, subsequently informs the generative task of the decoder by providing insights into the characteristics of the masked object. After the initial pre-training phase, the encoder is specifically fine-tuned for synthesis tasks, enhancing its adaptability to synthesis.

\subsubsection{Conditional DiT} Our objective is to model the distribution of the unmasked region $x_0^m$ conditioned on the masked region $x_0^v$ as $p(x_0^m|x_0^v)$.

\textbf{Forward diffusion process.} During the forward diffusion process, only the unmasked area $x_0^m$ undergoes diffusion. This process involves the gradual addition of Gaussian noise over $T$ steps to the masked components, producing a sequence of states $x_1^m, x_2^m, \dots, x_T^m$. Each step follows a Markov chain, detailed below:
\begin{equation}
 \mathbb{P} ( x_{t}^{m} | x_{t-1}^{m} ) := \mathcal{N} ( x_{t}^{m}; \sqrt{1- \beta _{t}} x_{t-1}^{m}, \beta _{t} I) 
\end{equation}
where $I$ denotes the standard normal distribution. The $a_{t} := 1- \beta _{t}$ and  $ \bar{a_{t}}= \prod_{s=1}^{t}a_{s}$ are used, the forward process admits sampling $x_{t}^{m}$ at an arbitrary timestep $t$ outlined below: 
\begin{equation}
 \mathbb{P} ( x_{t}^{m} | X_{0}^{m} ) = \mathcal{N}  ( x_{t}^{m};  {\sqrt{\bar{a_{t}}}} x_{0}^{m}, (1-\bar{a_{t}}) I)  \\
\end{equation}
The variance schedule ensures that $\bar{a}_T$ at the final timestep $T$ is sufficiently small, enabling $\mathbb{P}(x_T^m)$ to closely resemble the standard normal distribution $N(0, I)$. This resemblance effectively sets the stage for initiating the reverse diffusion process.

\textbf{Reserve diffusion process.} For each timestep of the reverse diffusion process, given $x_t^m$ and the corresponding conditional $x_t^v$, denoising is performed on the distribution $p(x_0^m|x_0^v)$. This process is approximated by recursively sampling from $p(x_{t-1}^m|x_t^m, x_t^v)$, beginning with $x_T^m \sim N(0, I)$.

\begin{equation}
Q (x_{t-1}^{m}\mid x_{t}^{m} ,x_{t}^{v}) := \mathcal{N}  ( x_{t-1}^{m};  \mu _{\theta}(x_{t}^{m}, t,x_{t}^{v}),  \sigma _{\theta} (x_{t}^{m}, t,x_{t}^{v}) I ) \\
\end{equation}
where $\sigma _{\theta}$ is the variance of conditional distribution $P ( x_{t-1} ^ {m} \mid x_{t}^{m} , x_{t}^{v})$.

The PHMDiff is trained to synthesize the target modality by predicting the involved noise $\epsilon_{\theta }$ under the guidance of the $C_{t}^{m}$, which is formulated below:
\begin{equation}
     L _{\epsilon} = E_{x_{t}, \epsilon \sim N(0,I),t}  \left\| \epsilon - \epsilon _{\theta} (x_{t}^{m},t,x_{t}^{v})\right\|_{2}^{2} \\
\end{equation}

\subsubsection{Cross-Granularity Regularization} To further enhance synthetic performance, we employ Maximum-Mean Discrepancy (MMD) regularization to model the mutual information across different granularity levels, thus implementing Cross-Granularity Regularization (CGR). MMD quantifies the similarity between two distributions by comparing all their moments \cite{li2015generative}. Specifically, within the PHMDiff framework, we model the mutual information between sampled noise distributions and Gaussian distributions at three distinct resolutions—low, middle, and high. The granularity regularization loss for the lowest layer is defined as: 
\begin{equation}
\begin{split}
     L_{l}(\epsilon \parallel m)=\mathbb{K}(\epsilon,\epsilon^{'})-2\mathbb{K}(m,\epsilon)+\mathbb{K}(m,m^{'})\\
     m=\epsilon _{\theta}(p(x),\sqrt{\bar{a_{t}}}x_{0}+\sqrt{1-\bar{a_{t}}\epsilon}+(1-\sqrt{\bar{a_{t}}})\hat{Y})
\end{split}
\end{equation}
where $\epsilon$ represents the noise, and $\mathbb{K}$ is a positive definite kernel used to reproduce distributions in the Hilbert space. CGR not only preserves the mutual information between the synthesis and priors at each granularity level but also ensures pixel-level detail and overall structural consistency across the hierarchical pyramid structure. Consequently, the combined loss, which includes Cross-Granularity losses ($L_{l}$, $L_{m}$, and $L_{h}$) and $L_{\epsilon}$, effectively optimizes network performance by synergistically enhancing both local and global features.

\section{Experiments}
\subsection{Datasets} We demonstrated the proposed PHMDiff model on two widely used multi-modality datasets: the pelvic MRI-CT dataset \cite{nyholm2018mr} and the BraTS 2021 dataset\footnote{http://braintumorsegmentation.org/}. The pelvic dataset comprises T2-weighted MR ($512 \times 512$) and CT ($512 \times 512$) images of the male pelvis from 15 subjects, with a split of 9 for training, 2 for validation, and 4 for testing. Each subject provided 90 axial cross-sections. The dataset, collected using various protocols and scanners, includes multi-modal images co-registered to T2-weighted MR scans, enhancing their utility for diverse research applications. And, the Brain Tumor Segmentation Challenge 2021 (BraTS 2021) \cite{baid2021rsna,menze2014multimodal,bakas2017advancing} includes 1,251 cases, each featuring four MRI sequences: T1, T2, FLAIR, and T1ce. These images were sourced from multiple institutions using varying protocols and scanners. A standardized pre-processing regimen was uniformly applied across all sequences to ensure consistency. This involved resampling the dimensions of each dataset to $240 \times 240 \times 150$ and normalizing the intensity values to a range of $\left[-1,1\right]$.

\subsection{Implementation details} 
We evaluated the performance of our network and other methods on the two datasets using PSNR (Peak Signal-to-Noise Ratio) and SSIM (Structural Similarity Index). The significance of performance differences was assessed with a paired t-test, with a threshold of p $<$ 0.05. A 5-fold cross-validation approach was employed to evaluate and compare the network's performance. The network was implemented on an Ubuntu 18.04 platform, using Python 3.8 and PyTorch 1.8. All computations were conducted on two NVIDIA RTX 3090 GPUs, each equipped with 24 GB of memory. The networks were trained using the Adam optimizer, with an initial learning rate set to $10^{-6}$ and a mini-batch size of 10. We set the scaling factor $\alpha$ to 0.5 in Eq.\ref{eq1}.

\subsection{Experimental Results}
\begin{figure*}[t]
\center
\includegraphics[width=0.9\textwidth]{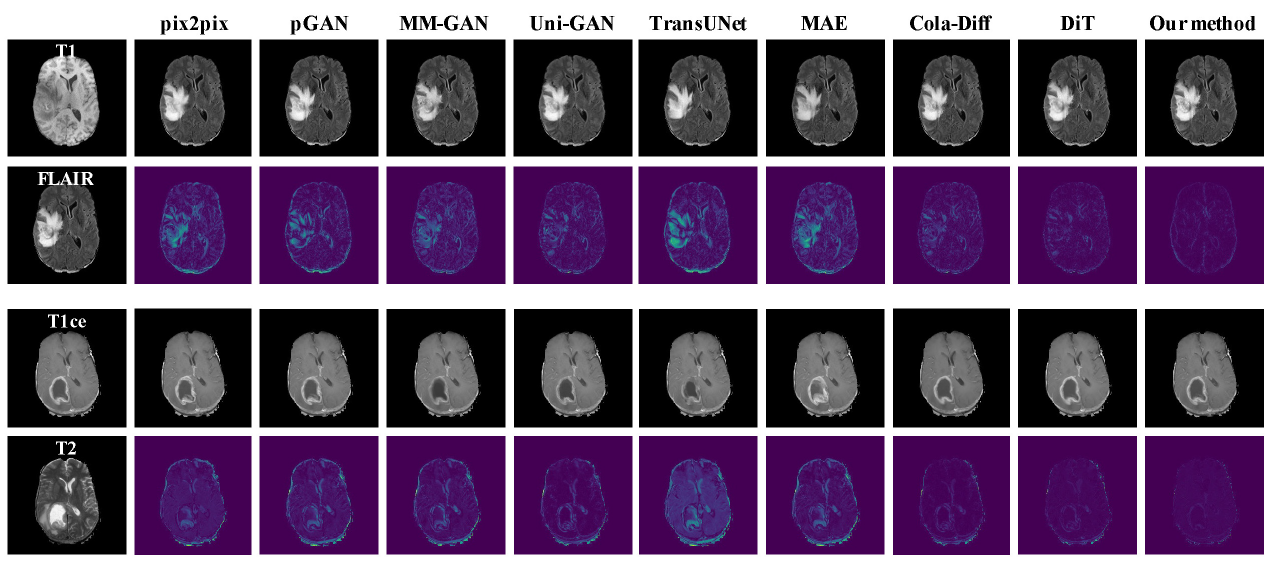}
\caption{Illustrative instances of synthetic images were demonstrated on the BraTS dataset for T1 $\rightarrow$FLAIR and T1ce$\rightarrow$T2. Synthesized images from all competing methods and the source and reference target modality are shown. Compared with the SOTA method, our method synthesizes images with lower noise and clearer texture details, edges, and shapes.} \label{SynBra}
\end{figure*}

\subsubsection{Comparison Settings} To demonstrate the superiority of our proposed framework, PHMDiff was compared with several baseline methods across two datasets. The baseline methods include GAN-based models (pix2pix \cite{isola2017image}, pGAN \cite{dar2019image}, MM-GAN \cite{sharma2019missing}, and Uni-GAN \cite{zhang2023unified}), a Transformer-based model (TransUNet \cite{chen2021transunet}), MAE, and Diffusion-based models (DiT\cite{peebles2023scalable} and Cola-Diff \cite{jiang2023cola}). It is important to note that Cola-Diff uses all other available modalities as conditions. To ensure fairness in the experiments, a consistent approach was used with LDM \cite{rombach2022high} when only a single modality was input. The hyperparameters for each competing method were optimized using identical cross-validation procedures.

\subsubsection{Synthesis Results Comparison with SOTA} To quantitatively evaluate the synthesis performance of our method, we compared the performance of our proposed PHMDiff model with existing state-of-the-art synthesis methods on the BraTS dataset for the tasks T1$\rightarrow$T2 and FLAIR$\rightarrow$T1. The performance metrics employed include PSNR and SSIM. The statistical analysis results of the p-Values ($<$ 0.05) show that the difference between the proposed method and each competing method is significant. As shown in Table \ref{Brats_sota}, PHMDiff demonstrated superior performance, achieving a PSNR of 28.32$\pm$1.16 dB and an SSIM of 92.42$\pm$1.53\% for T1$\rightarrow$T2, and a PSNR of 27.95$\pm$1.27 dB with an SSIM of 92.15$\pm$1.48\% for FLAIR$\rightarrow$T1. The results of statistical analysis of the p-Values ($<$ 0.05) via paired t-test show that the difference between our PHMDiff and the other related methods is significant. 

\begin{table}[t]
\caption{Quantitative comparison with synthesis methods in various one-to-one tasks on BraTS dataset (T1$\rightarrow$T2 and FLAIR$\rightarrow$T1. PSNR($dB$) and SSIM(\%) are listed and reported values are mean $\pm$ std (\textbf{orange} indicates the top-performing model).} \label{Brats_sota}
\begin{adjustbox}{width=1\linewidth}
\begin{tabular}{lllll}
\hline
\multirow{2}{*}{Model} & \multicolumn{2}{l}{T1$\rightarrow$T2} & \multicolumn{2}{l}{FLAIR$\rightarrow$T1} \\
 & PSNR & SSIM & PSNR & SSIM \\ \hline
pix2pix & $22.73_{\pm1.23}$ & $84.15_{\pm1.44}$ & $22.39_{\pm0.93}$ & $83.78_{\pm1.56}$ \\
pGAN & $24.59_{\pm1.34}$ & $85.52_{\pm1.69}$ & $24.04_{\pm1.14}$ & $85.17_{\pm1.24}$ \\
MM-GAN & $24.87_{\pm1.15}$ & $85.66_{\pm1.24}$ & $24.57_{\pm1.39}$ & $85.23_{\pm1.37}$ \\
Uni-GAN & \textcolor{blue}{\uline{$26.46_{\pm1.47}$}} & \textcolor{blue}{\uline{$87.31_{\pm1.15}$}} & \textcolor{blue}{\uline{$26.12_{\pm1.25}$}} & \textcolor{blue}{\uline{$87.04_{\pm0.93}$}} \\
TransUNet & $21.35_{\pm1.59}$ & $83.13_{\pm0.98}$ & $20.92_{\pm1.17}$ & $81.42_{\pm1.06}$ \\
MAE & $20.41_{\pm1.03}$ & $78.74_{\pm1.12}$ & $20.29_{\pm1.26}$ & $78.31_{\pm1.37}$ \\
Cola-Diff & $25.76_{\pm0.96}$ & $86.54_{\pm0.93}$ & $25.33_{\pm0.83}$ & $86.26_{\pm1.54}$ \\
DiT & $25.97_{\pm1.27}$ & $86.89_{\pm1.26}$ & $25.51_{\pm1.34}$ & $86.36_{\pm1.39}$ \\
PHMDiff & \cellcolor{orange!50}$28.32_{\pm1.16}$ & \cellcolor{orange!50}$92.42_{\pm1.53}$ & \cellcolor{orange!50}$27.95_{\pm1.27}$ & \cellcolor{orange!50}$92.15_{\pm1.48}$ \\ \hline
\end{tabular}

\end{adjustbox}
\end{table}

Fig.\ref{SynBra} presents a comparative visualization of synthetic results generated by various state-of-the-art methods for the T1$\rightarrow$FLAIR and T1ce$\rightarrow$T2 synthesis tasks on the BraTS dataset, highlighting that PHMDiff achieves the best synthesis performance among the methods evaluated. The figure includes synthesized images alongside the original MRI and the target modality, offering a visual assessment of each method's ability to replicate the target MRI sequence accurately. In the synthesized images, PHMDiff notably improves the preservation of complex anatomical structures, particularly at challenging boundaries and within detailed textures. The error maps included in our analysis further underscore the areas where PHMDiff excels in maintaining crucial boundaries and textures more effectively than competing methods. These maps illustrate synthesis accuracy differences, with PHMDiff showing fewer discrepancies from the target modalities, underscoring its superior performance. The effectiveness of PHMDiff can be attributed to its innovative hierarchical diffusion process, which adeptly manages multi-scale information. This process ensures that both high-level anatomical features and fine details are preserved, dynamically adapting to the complexity of each image region. Additionally, PHMDiff incorporates a robust regularization strategy that maintains consistency across various levels of detail and resolution. The alignment between our quantitative and qualitative findings further validates the superior synthesis performance of PHMDiff.

\subsubsection{Synthesis Results Comparison with SOTA on the Pelvic Dataset} 
\begin{figure}[t]
\center
\includegraphics[width=0.45\textwidth]{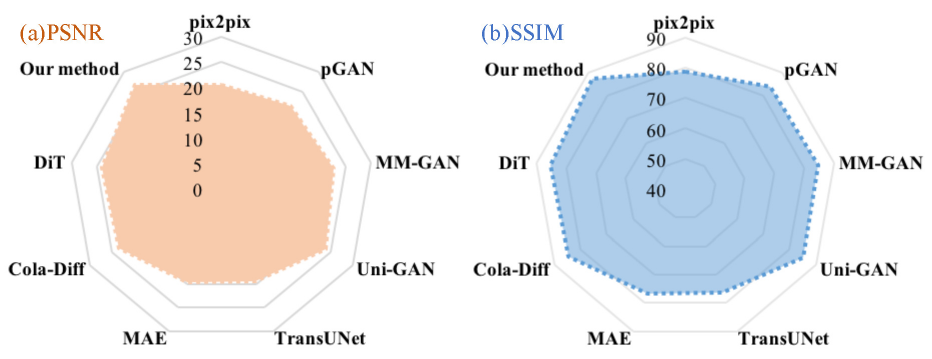}
\caption{Quantitative comparison with other synthesis methods in MRI $\rightarrow$ CT tasks on Pelvic dataset. The experimental results of our method compared with other SOTA methods in terms of (a) PSNR and (b) SSIM.} \label{MRI}
\end{figure}

To quantitatively evaluate the synthesis performance of our method, we compared the performance of our proposed PHMDiff model with existing state-of-the-art synthesis methods on the Pelvic dataset for the cross-modality task of MRI$\rightarrow$CT. As illustrated in the radar chart (Fig. \ref{MRI}), PHMDiff achieves significantly higher PSNR and SSIM scores. Specifically, the PHMDiff curve encompasses a larger area than other methods, indicating superior performance across all metrics. This superior performance demonstrates that PHMDiff synthesizes images with better quality and greater structural similarity to the target modality. 

\begin{figure}[t]
\center
\includegraphics[width=0.46\textwidth]{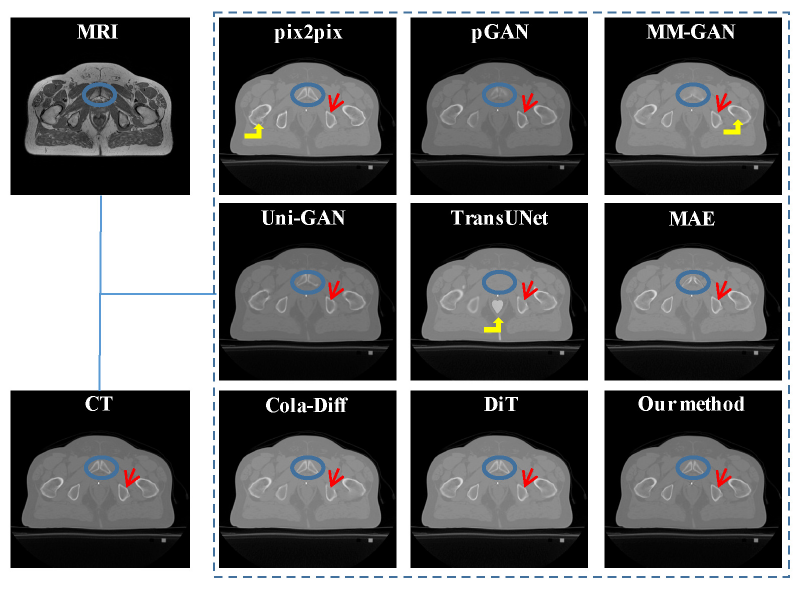}
\caption{Illustrative instances of synthetic images on the Pelvic dataset for MRI $\rightarrow$ CT. Compared with the SOTA methods, our method synthesizes images with lower noise and clearer texture details, edges, and shapes.} \label{SynPelvic}
\end{figure}

The visualized comparison results are presented in Fig.\ref{SynPelvic}, showcasing representative MRI$\rightarrow$CT synthesis tasks on the Pelvic dataset. These results indicate that PHMDiff achieves the best-synthesized performance compared to other state-of-the-art (SOTA) methods. Our method produces target images with reduced noise and more precise textural and edge definitions compared to baseline models. Specifically, the blue elliptical region in the figure highlights significant discrepancies in synthesis quality across different models. In the MRI modality, this area features weaker boundaries between adjacent tissues and organs, making it susceptible to loss during the synthesis process. Notably, in the images synthesized by TransUNet, the delineation of this region is almost entirely lost. In contrast, PHMDiff excels at preserving the integrity of these boundaries, as indicated by the yellow and red arrows in the figure, which point to specific boundaries that closely match the ground truth CT images. This outcome underscores the efficacy of PHMDiff in capturing critical structural details that are often compromised in other synthesis methods. Furthermore, the consistency between our quantitative and qualitative findings further validates the superior synthesis performance of the PHMDiff.

\subsubsection{Ablation study}

To assess the individual contributions of components to the synthesis process, we conducted a comparative analysis featuring our complete PHMDiff model alongside variants lacking each of these components: without the pyramid hierarchical structure (w/o PH), w/o MAE, w/o Diff, w/o Transformer, and w/o CGR. The findings from this comparison underscore the superior performance of our integrated PHMDiff model, which consistently outperformed the component-specific variants. As detailed in the ablation study results presented in Table.\ref{ablation}, the complete PHMDiff configuration achieved the highest scores for both PSNR and SSIM, affirming the enhanced image quality and structural integrity of the synthesized images produced by our full model. The absence of any single component generally led to a decline in performance. Specifically, removing the cross-granularity regularization significantly impacted the model’s ability to maintain consistency across varying levels of detail, which is crucial for achieving accurate pixel-level perceptual quality. Similarly, excluding the Transformer component reduced the model's capability to effectively capture global dependencies and contextual nuances essential for accurately synthesizing images across different regions. The elimination of either the diffusion component or MAE resulted in lower scores, highlighting their critical roles in enhancing the synthesis process and overall fidelity of the generated images. These results not only validate the essential contributions of each component to the PHMDiff model but also demonstrate the benefits of integrating a pyramid hierarchical structure to more effectively manage the synthesis process, thereby significantly improving performance metrics compared to baseline models.

\begin{table}[t]
\caption{Quantitative comparison with ablation study in various one-to-one tasks on BraTS dataset (T1$\rightarrow$T2 and T1$\rightarrow$T1ce. PSNR($dB$) and SSIM(\%) are listed and reported values are mean $\pm$ std. The \textbf{orange} indicates the top-performing model.} \label{ablation}
\begin{adjustbox}{width=1\linewidth}
\begin{tabular}{lllll}
\hline
\multirow{2}{*}{Model} & \multicolumn{2}{l}{T1$\rightarrow$T1ce} & \multicolumn{2}{l}{T1$\rightarrow$T1ce} \\
                       & PSNR        & SSIM        & PSNR          & SSIM         \\\hline
w/o CGR & \textcolor{blue}{\uline{$27.83_{\pm1.23}$}} & \textcolor{blue}{\uline{$91.77_{\pm0.93}$}} & \textcolor{blue}{\uline{$28.67_{\pm0.87}$}} & \textcolor{blue}{\uline{$91.95_{\pm1.38}$}} \\
w/o Transformer & $23.48_{\pm0.89}$ & $86.59_{\pm1.01}$ & $24.13_{\pm1.25}$ & $87.68_{\pm1.25}$ \\
w/o Diff & $26.54_{\pm0.96}$ & $89.93_{\pm1.16}$ & $27.78_{\pm0.95}$ & $90.39_{\pm1.66}$ \\
w/o MAE & $26.46_{\pm1.47}$ & $87.31_{\pm1.15}$ & $26.12_{\pm1.25}$ & $87.04_{\pm0.93}$ \\
w/o PH & $25.98_{\pm1.45}$ & $88.86_{\pm1.37}$ & $26.64_{\pm0.79}$ & $89.03_{\pm1.42}$ \\
PHMDiff & \cellcolor{orange!50}$28.32_{\pm1.16}$ & \cellcolor{orange!50}$92.42_{\pm1.53}$ & \cellcolor{orange!50}$29.49_{\pm1.34}$ & \cellcolor{orange!50}$93.58_{\pm0.87}$ \\ \hline
\end{tabular}
\end{adjustbox}
\end{table}

\begin{figure}[t]
\center
\includegraphics[width=0.48\textwidth]{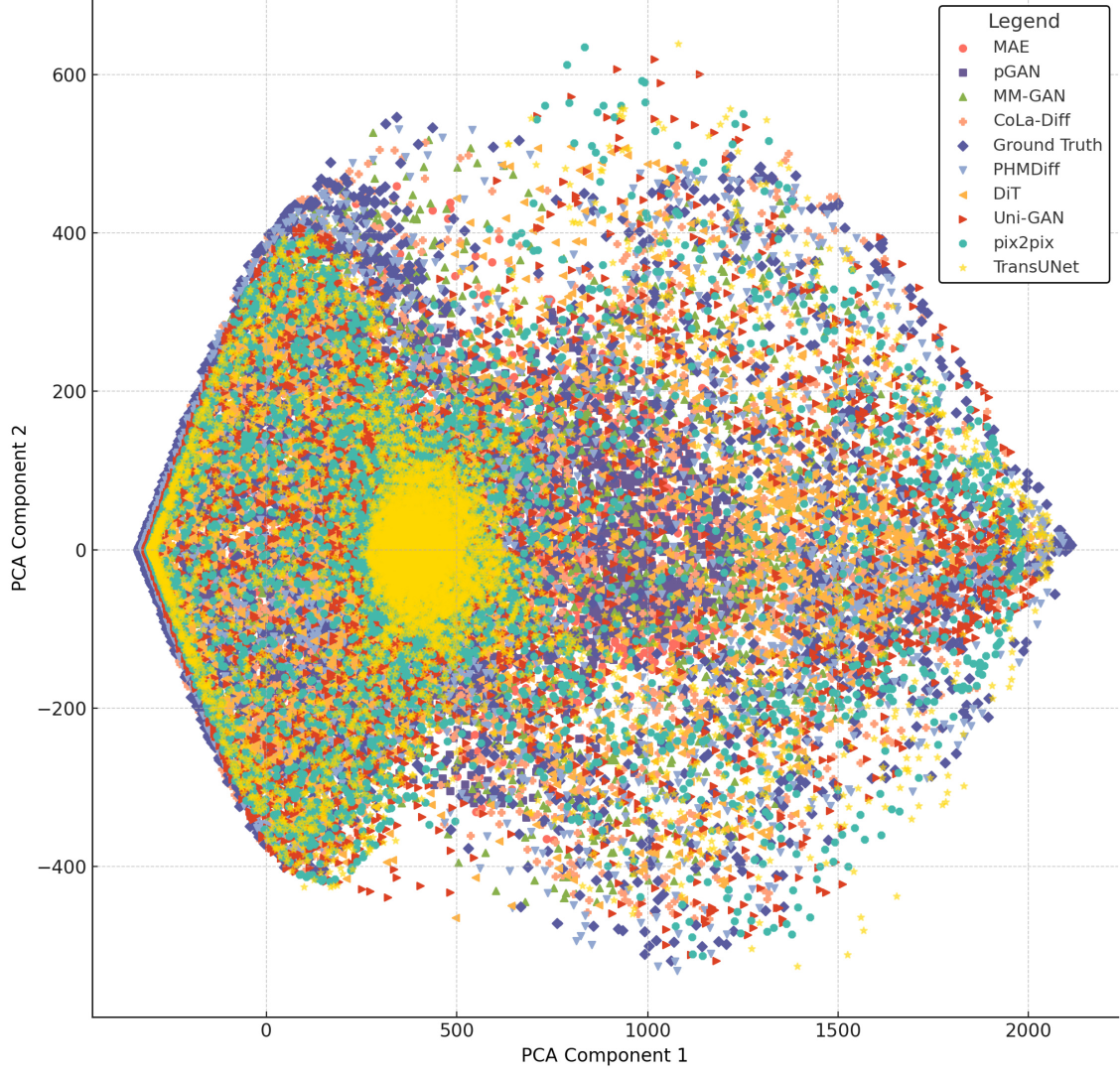}
\caption{The t-SNE feature space visualization for the different model's synthetic images.} \label{pca}
\end{figure}

\subsubsection{PHMDiff's Promotion of Synthesis} Fig.\ref{pca} presents the t-SNE visualization \cite{van2008visualizing} of the image patch feature space for synthetic images generated by various models. Each point in the scatter plot represents a 3$\times$3 patch of the original image, projected onto the first two principal components using principal component analysis (PCA). The feature representations for different methods-ground truth (indigo blue), pix2pix (teal), pGAN (deep purple), MM-GAN (olive green), Uni-GAN (red-orange), TransUNet (gold), MAE (light coral), CoLa-Diff (light salmon), DiT (light orange), and our method (light blue)- are shown for brain MRI. Significant overlap of colors indicates that the synthesized images share similar anatomical structures, contrast levels, or other common features with the ground truth slices. However, the spread of different color components across the plot indicates variability in feature representation among the images generated by different models. A broader spread of certain colors, such as yellow for MAE, suggests a greater deviation from the real images.

\begin{figure}[t]
\center
\includegraphics[width=0.42\textwidth]{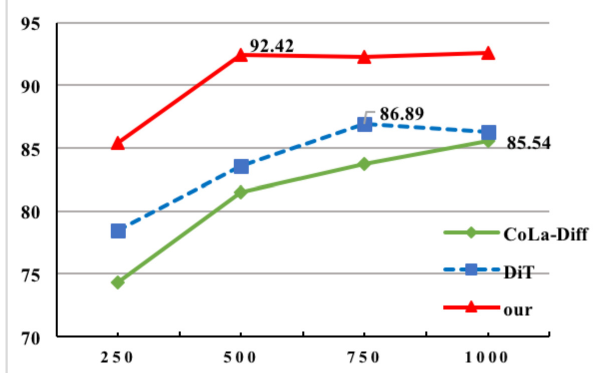}
\caption{experimental results of our PHMDiff compared with DiT and CoLa-Diff at different time steps for SSIM on BraTS dataset (T1$\rightarrow$T2).} \label{speed}
\end{figure}

As the number of timesteps increases, the quality of the synthesized images improves significantly. Consider Fig.\ref{speed}, our PHMDiff model, trained with only 500 timesteps, surpasses the performance of both DiT and CoLa-Diff models trained with 1000 timesteps. This demonstrates that the pyramid hierarchical coarse-to-fine synthesis, multi-scale random masking strategy, and the Transformer's ability to capture long-range dependencies enable our approach to achieve or even exceed the performance of other diffusion-based methods with fewer training steps, thereby reducing the overall training cost. 

\subsection{Impact of Pyramid Structure.} 
\begin{figure}[t]
\center
\includegraphics[width=0.48\textwidth]{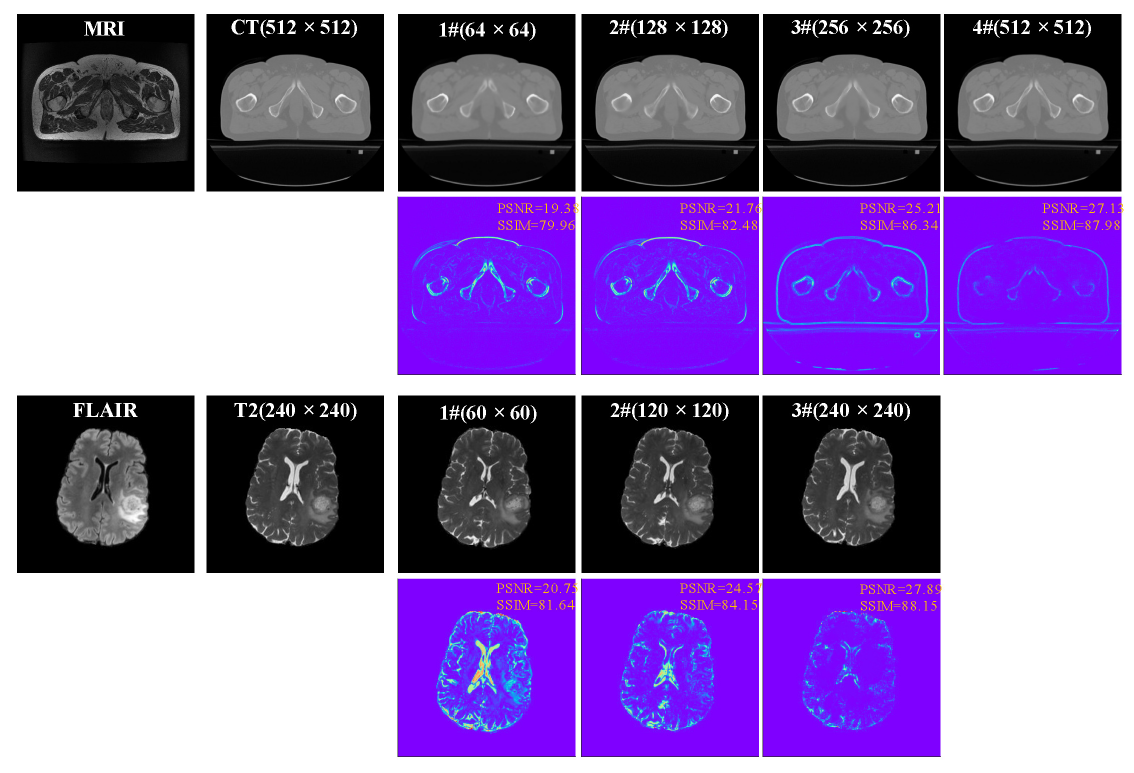}
\caption{Illustrative instances of synthetic images were demonstrated on the MRI$\rightarrow$CT task on the Pelvic dataset and the FLAIR$\rightarrow$T2 task on the BraTS dataset. Synthesized images from all competing methods are shown along with the source and reference target images. Error plots can more intuitively observe the differences between the synthesized image and the ground truth, thereby reflecting the quality of the synthesis.} \label{layer}
\end{figure}

To validate the effectiveness of the proposed pyramid structure, we visualized the synthetic outcomes at each layer. As depicted in Fig.\ref{layer}, visualizations for the MRI$\rightarrow$CT task on the Pelvic dataset and the FLAIR$\rightarrow$T2 task on the BraTS dataset are presented. Error maps provide a clear visual indication of the discrepancies between the synthetic results and the ground truth, substantiating that the coarse-to-fine synthesis progresses to fine resolution with each layer, enhancing both global structures and local details. Additionally, both PSNR and SSIM metrics show incremental improvements with each successive layer. These outcomes confirm that such a multi-scale pyramid structure can effectively accelerate and enhance the quality of synthesis.

\subsection{Impact of Synthetic Data on Task Performance}

We conduct segmentation using real data, synthetic data, and a combination of both ('All') across different segmentation frameworks. Among these, LF-SynthSeg \cite{xu2024lf} is a unified framework specifically designed for brain tumor synthesis and segmentation. As shown in Table.\ref{segmentation}, the 'All' dataset demonstrates superior performance compared to the other datasets. This outcome provides strong evidence that incorporating synthetic data via our proposed PHMDiff approach can significantly enhance segmentation accuracy. Such improvements underscore the utility of synthetic data in enriching training datasets and augmenting model robustness, ultimately leading to more precise and reliable medical image analysis.

\begin{table}[t]
\caption{Segmentation of Brain MRI using UNet and nnUNet with DSC and HD95 metrics (mean $\pm$ std).} \label{segmentation}
\centering
\begin{tabular}{llll}
\hline
 &  & DSC & HD95 \\ \hline
\multirow{3}{*}{UNet} & Syn & $0.82_{\pm0.13}$ & $17.46_{\pm7.39}$ \\
 & Real & $0.85_{\pm0.09}$ & $13.74_{\pm10.45}$ \\
 & All & $\textcolor{blue}{\uline0.88_{\pm0.07}}$ & $\textcolor{blue}{\uline11.02_{\pm8.87}}$ \\ \hline
\multirow{3}{*}{nnUNet} & Syn & $0.83_{\pm0.12}$ & $15.93_{\pm9.26}$ \\
 & Real & $0.87_{\pm0.06}$ & $11.49_{\pm12.34}$ \\
 & All & \cellcolor{orange!50}$0.92_{\pm0.03}$ & \cellcolor{orange!50}$6.37_{\pm5.39}$ \\\hline
LF-SynthSeg & Syn & $0.80_{\pm0.11}$ & $15.66_{\pm16.62}$ \\ \hline
\end{tabular}
\end{table}

\section{Conclusion}
In this paper, we introduce the Pyramid Hierarchical Masked Diffusion Model (PHMDiff), a novel network that combines Masked Autoencoders (MAE) with a Transformer-based Diffusion model for both cross-modal and intra-modal synthesis. The network employs a multi-scale pyramid structure for controlled, detail-oriented synthesis and uses multi-scale masks to enhance critical areas, improving image quality. Cross-granularity regularization ensures spatial consistency by integrating global and local information, optimizing detail and structural coherence. Our extensive experiments show that PHMDiff significantly outperforms existing methods, achieving superior PSNR and SSIM scores and producing high-quality images, thus demonstrating its potential impact in the field.



\end{document}